\documentstyle[preprint,aps]{revtex}
\input epsf.tex
\def\DESepsf(#1 width #2){\epsfxsize=#2 \epsfbox{#1}}
\begin{document}
\preprint{\vbox{\hbox{}}}
\draft
\title{
Gravitational Lensing And Extra Dimensions}
\author{Xiao-Gang He$^{1,2}$, 
Girish C. Joshi$^2$ and Bruce H.J. McKellar$^2$}
\address{
$^1$Department of Physics, National Taiwan University, Taipei, Taiwan 10617, R.O.C.\\
and\\
$^2$School of Physics, University of Melbourne, Parkville, Vic. 3052, Australia}

\date{August 1999}
\maketitle
\begin{abstract}
We study gravitational lensing and the bending of light 
in low energy scale ($M_S$)
gravity theories with extra space-time dimensions $n$.
We find that 
due to the presence of spin-2 Kaluza-Klein states from compactification, a
correction to the deflection angle with a strong quadratic
dependence on the photon energy is introduced. No deviation
from the Einstein General Relativity 
prediction for the deflection angle for photons grazing the Sun 
in the visible band with 15\% accuracy (90\% c.l.) implies that
the scale $M_S$ has to be larger than $1.4(2/(n-2))^{1/4}$ TeV and
approximately 4 TeV for n=2.
This lower bound is comparable 
with that from collider physics constraints. 
Gravitational lensing 
experiments with higher energy photons can provide stronger constraints.
\end{abstract}
\pacs{PACS Numbers: 04.50.th, 04.80.Co, 11.10.Kk, 12.60.-i}
\newpage

Gravitational lensing or the gravitational bending of light, 
is one of the most important evidence which supports the
Einstein General Relativity (EGR) theory\cite{1a}.
Light sources are deflected when passing by 
a massive object. In 
EGR theory at grazing incidence the deflection angle is
predicted to be
$\theta = 4G_N m/R$, where $m$ is the mass and $R$ the radius of the massive object. 
For the Sun the deflection angle is $1.75''$.
This prediction provides an important test for different theories of gravity\cite{1,2,3,3a}.
In fact the detection of deflection angle of light passing by the Sun in 1919 
was one of the most important first experiments which supported EGR theory\cite{4a}.  
Since then many other experiments have been carried out and 
found no deviation from the EGR theory\cite{4,5,radio,visible}.
It is usual to measure deviation from the EGR theory 
in terms of the post-Newtonian
parameter $\gamma$ defined by $\theta = (4G_N m/R)(1+\gamma)/2$ which is one in EGR theory.
The EGR theory is in agreement within a level better than one 
percent with experiments in the 
radio band to visible band\cite{4,5}. 

There are other alternative theories for gravity, such as
tensor-scalar theories\cite{2} or theories with extra dimensions\cite{3,3a}.
It is important to establish to what extent these theories are consistent 
with experiments in order
to find the ultimate theory of gravity.
In these alternative theories due to different type of gravitational 
interaction or new interactions in
addition to the standard EGR interaction, there will be corrections to the parameter
$\gamma$. Experimental measurements thus can provide strong constraints for other theories or
even rule out some theories\cite{1,2}. 
In this paper we study gravitational lensing in theories with
extra space-time dimensions.

It has recently been proposed
that gravitational effects can become large at a scale
$M_S$ near the weak scale due to effects from extra dimensions\cite{3,3a},
which is 
quite different from the traditional concept that gravitational effects
only become large at the Planck scale $M_{Pl}=
\sqrt{1/G_N} \sim 10^{19}$ GeV.
In this proposal the total space-time has $D=4 + n$ dimensions.
The relation between the scale $M_S$ and the Planck scale
$M_{Pl}$, assuming all extra dimensions are compactified with the
same size R, is given by $M^2_{Pl} \sim R^n M^{2+n}_S$.  
For $n\geq 2$, $M_S$ can be of order one TeV and $R$ can be in the
sub-millimeter region\cite{3}. 
When the extra dimensions are compactified
there are towers of
states, the Kaluza-Klein (KK) states with spin-2, spin-1 and spin-0, 
which interact with ordinary matter fields.
There are many interesting consequences for collider physics\cite{6,7}, 
astrophysics
and cosmology\cite{8}.
These new interactions provide information about the allowed value for
$M_S$.
The lower bound for $M_S$ is
constrained, typically, to be of order one TeV from collider 
experimental data\cite{6,7}. There are also constraints from
cosmological and astrophyiscal considerations\cite{8}.

Gravitational lensing is due to exchange of a massless 
graviton between photons and massive 
objects in EGR theory. In theories with extra dimensions gravitational lensing also 
receives contributions from the
massive KK states in addition to that from the usual one.
The massive KK states couple to matter fields in a way similar to the massless graviton.
This makes gravitational lensing a sensitive test of theories with 
extra dimensions.
We indeed find that the effects from 
massive KK states are significant and a term strongly dependent on the  photon energy 
is introduced in the expression for the deflection angle
 if the scale $M_S$ is in the TeV region. Experimental 
data on gravitational 
lensing by the Sun can provide interesting bounds on the scale
$M_S$ for these theories. The observation that there is 
no deviation from the ERG prediction with $\gamma -1< 15\%$ (90\% confidence 
level) in 
the range of visible light for light at grazing incidence to the Sun 
requires $M_S$ to be larger than
$1.4(2/(n-2))^{1/4}$ TeV and approximately 4 TeV for n=2. 
This bound is comparable to 
that from collider physics experiments\cite{6,7}. 
Gravitational lensing experiments with higher energies 
are able to put even more stringent limits on $M_S$. 
For a $\gamma$-ray of energy one MeV, 
no observed 
deviation from ERG at the 10\% level would 
set a lower limit of $1.5\times 10^3$ TeV
for $M_S$. 

After compactifying the extra n dimensions, for a given KK level
$\vec l$ there are one spin-2, n-1 spin-1 and n(n-1)/2
spin-0 states\cite{7}. Assuming that all standard fields are confined to
a four dimensional world-volume and gravitation is minimally
coupled to standard fields, it was found that the spin-1 KK states
decouple while the spin-2 and spin-0 KK states couple to all standard 
fields\cite{7}.
We, however, found that only spin-2 KK states can interact with both the photon and the Sun.
The graviton and the spin-2 KK states couple to the 
energy momentum tensor of the Sun which is similar to the coupling of a spin-2 
particle to 
a scalar. There are different ways to obtain the deflection angle of light by 
a massive object.  We will 
treat the Sun as a scalar S and obtain the deflection 
angle by matching the scattering cross section and the impact 
parameter.
The process studied is similar to photon-Higgs scattering\cite{9}.
The Feynman diagram is shown in Figure 1. Using the
Feynman rules given in Ref.\cite{7}, we obtain the scattering amplitude for, 
 $\gamma(\epsilon_1(p_1))
+S(k_1) \to \gamma(\epsilon_2(p_2)) + S(k_2)$, as

\begin{eqnarray}
M
&=&- {4\pi G_N} (m^2\eta^{\mu\nu} +
C^{\mu\nu,\rho\sigma} k_{1\rho} k_{2\sigma})
\left (
{B^{graviton}_{\mu\nu,\alpha\beta}\over q^2}+
\sum_l {B^{KK}_{\mu\nu,\alpha\beta}\over q^2-m_l^2}\right )\nonumber\\
&\times& (p_1\cdot p_2 C^{\alpha\beta,\delta\gamma}
+ D^{\alpha\beta,\delta\gamma}) \epsilon_{1\delta}(p_1)
 \epsilon^*_{2\gamma}(p_2),\nonumber\\
C^{\mu\nu,\rho\sigma} &=& \eta^{\mu\rho}\eta^{\nu\sigma}
+\eta^{\mu\sigma}\eta^{\nu\rho}-\eta^{\mu\nu} \eta^{\rho\sigma},\nonumber\\
D^{\alpha\beta,\delta\gamma}
&=& \eta^{\alpha\beta} k_1^\delta k_2^\gamma
-[\eta^{\alpha\gamma} k_1^\beta k_2^\delta + \eta^{\alpha \delta}
k_1^\gamma k_2^\beta
-\eta^{\delta\gamma} k_1^\alpha k_2^\beta + (\alpha \to \beta,
\beta\to \alpha)],\nonumber\\
B^{graviton}_{\mu\nu,\alpha\beta} &=&\eta_{\mu\alpha}\eta_{\nu\beta}
+\eta_{\nu\alpha}\eta_{\mu\beta} - \eta_{\mu\nu} \eta_{\alpha\beta},\nonumber\\
B^{KK}_{\mu\nu,\alpha\beta}
&=& (\eta_{\mu\alpha}- {q_\mu q_\alpha\over m_l^2})
(\eta_{\nu\beta}-{q_\nu q_\beta
\over m_l^2})+(\eta_{\nu\alpha} - {q_\nu q_\alpha\over m^2_l})
(\eta_{\mu\beta}-{q_\mu q_\beta\over m_l^2})\nonumber\\
&-&{2\over 3} (\eta_{\mu\nu} -
{q_\mu q_\nu\over m_l^2})
(\eta_{\alpha\beta} - {q_\alpha q_\beta\over m_l^2}).
\end{eqnarray}
where $q^2 = (p_1-p_2)^2$, m is the scalar mass, and $m_{l}$ is the mass of KK
state.
The sum is over all possible massive KK states.
The term proportional to $B^{graviton}_{\mu\nu,\alpha\beta}$ is
the contribution from EGR theory due to the massless graviton, and the term
proportional to $B^{KK}_{\mu\nu,\alpha\beta}$ is the contribution
due to the KK states. Gauge invariance dictates that the contributions from 
terms 
in $B^{gravity}_{\mu\nu,\alpha\beta}$ and $B^{KK}_{\mu\nu,\alpha\beta}$ 
proportional to
$\eta_{\mu\nu} \eta_{\alpha\beta}$ and any term which has an uncontracted
Lorentz index 
on $q$ vanish.
Due to this property, the total contribution 
is simply related the pure massless graviton one 
by replacing $1/q^2$ by
$1/q^2 + \sum_l 1/(q^2-m^2_l)$. 
We have

\begin{eqnarray}
M&=&-16G_N \left({1\over q^2} + \sum_l {1\over q^2-m_l^2}\right )
\left (
\epsilon_1\cdot \epsilon_2^*
[p_1\cdot k_1 p_2\cdot k_2 + p_2\cdot k_1 p_1\cdot k_2 - p_1\cdot p_2 k_1\cdot
k_2] \right .\nonumber\\
&+& p_1\cdot p_2 [\epsilon_1\cdot k_1 \epsilon^*_2\cdot k_2 +
\epsilon_1\cdot k_2 \epsilon^*_2\cdot k_1] 
+k_1\cdot k_2 \epsilon_1\cdot p_2 \epsilon_2^*\cdot p_1\nonumber\\
&-&
\left . p_1\cdot k_2 \epsilon_1\cdot p_2 \epsilon^*_2\cdot k_1
-p_1\cdot k_1 \epsilon_1\cdot p_2 \epsilon_2^*\cdot k_2
-p_2\cdot k_2 \epsilon_1\cdot k_1 \epsilon^*_2\cdot p_1 
-p_2\cdot k_1 \epsilon_1\cdot k_2 \epsilon^*_2\cdot p_1 \right ).
\end{eqnarray}

For small deflection angles the photon energies $\omega_1$ and $\omega_2$ 
are approximately the same which will be indicated by 
$\omega$, and $q^2 = -4\omega_1\omega_1 \sin^2(\tilde\theta/2)
\approx -\omega^2\tilde \theta^2$. Here $\tilde \theta$ is the angle between the 
incoming and outgoing photon directions.
Neglecting small terms proportional to $\tilde \theta$ 
in the numerator,
we obtain 

\begin{eqnarray}
{d\sigma\over d \Omega}
=16 G^2_N m^2 \left ( {1\over q^2}
+\sum_l {1\over q^2 - m^2_l} \right )^2.
\end{eqnarray}
Without the
massive KK contribution, the result reduces to the standard one. 

All possible KK
states have to be summed over. The masses for the KK states are given by
$m^2_l = 4\pi^2 \vec l^2/ R^2$, where $\vec l$ represents the hyper-cubic
lattice sites in n-dimensions. For $M_S$ in the multi-TeV range the
KK states are nearly degenerate and the sum can be approximated by integral in
n-dimensions.
Using the result in Ref.\cite{7},
we obtain

\begin{eqnarray}
\Delta = \sum_l {1\over q^2 - m^2_l}
=-{2\over M^4_S G_N} \left (|q^2|\over M^2_S\right )^{n/2-1} 
I_n(M_S/\sqrt{|q^2|}),
\end{eqnarray}
with

\begin{eqnarray}
I_n = \int^{M_S/\sqrt{|q^2|}}_{M_{min}/\sqrt{|q^2|}}
{y^{n-1}\over 1+y^2} dy,
\end{eqnarray}
where $M_{min}$ is the minimal KK state mass $2\pi/R$ which is
of order $10^{-3}$ eV
 for $R$ in the milli-meter range. 
The leading contribution to $\Delta$ for $|q^2|/M_S^2 << 1$, which is true in our
case, is equal to
$(1/(M_S^4 G_N)) \ln(M^2_S/(|q^2| + M^2_{min})$ for $n=2$, and
$(1/(M^4_S G_N))(2/(n-2))$ for $n>2$.
In the above we have
used $G_N = (4\pi)^{n/2}
\Gamma(n/2)R^{-n}M_S^{-(n+2)}$.

The differential cross section for small
scattering angle $\tilde \theta$ can be written as

\begin{eqnarray}
{d\sigma\over d\Omega}
= 16 G_N^2 m^2 \left ({1\over \tilde \theta^2} + \omega^2\Delta \right )^2,
\end{eqnarray}
Keeping the leading correction to the 
deflection angle $\theta$, we obtain
\begin{eqnarray}
\theta = {4G_N m\over R}
\left ( 1- 2\omega^2 \Delta \left ( {4G_N m\over R}\right )^2
\ln \left ({4G_N m\over R}\right ) \right ).
\end{eqnarray}
We note that effect of extra dimensions is always to increase the deflection angle, 
and also 
to introduce a $\omega$ dependence in the deflection angle. 
In the EGR theory an $\omega$ dependence can be generated 
at one loop order\cite{10}.
However, there the contribution is
extremely small. 
The contribution from extra dimensions obtained here can be very 
large--close to the present experimental reach. 
For easy comparison with data we 
work with the post-Newtonian
parameter $\gamma$.
The expression for $\theta$ gives the correction 
$\Delta \gamma = \gamma -1$ as

\begin{eqnarray}
\Delta \gamma = -4\omega^2 \Delta \left ({4G_N m\over
R}\right )^2 \ln \left ({4G_N m\over R}\right ).
\end{eqnarray}

For the Sun $R_\odot = 6.96\times 10^{5}$ km, and
$2G_N m_\odot = 2.95$ km, we obtain the correction to $\gamma$ for  
grazing deflection of light by the Sun 

\begin{eqnarray}
\Delta \gamma = -0.50 \left ({\omega^2\over \rm{eV}^2}\right )
\left ( {1\rm{TeV}\over M_S}\right )^4 \delta,
\end{eqnarray}
with $\delta = 2/(n-2)$ for $n>2$, and $\delta \approx \ln (M^2_S/(\omega^2 
\theta^2 +m^2_{min})$ for $n=2$. For n=2, $\delta$ is of order
50 to 80 for $\omega$ in the range of radio waves to $\gamma$-rays and $M_S$ 
in the range of multi-TeV. 
Using the above one can extract important information about theories
with extra dimensions.
 
Our calculations correspond to the determination of the total deflection of 
the light coming from a distant source and grazing the Sun, 
so that the impact parameter is $R_\odot$. This can easily be generalized to the total
deflection with an arbitrary impact parameter $b$, simply by replacing 
$R_\odot$ by $b$. As the correction induced by the KK states to the EGR 
is small, we may assume that the photon follows a post Newtonian 
geodesic path to 
obtain the deflection angle $\delta \alpha$ measured at the Earth\cite{wheeler} with

\begin{eqnarray}
\delta\alpha = {(1+\gamma)G_Nm_\odot\over r_E}
{\sin \alpha\over 1-\cos \alpha},
\end{eqnarray}
where $\alpha$ is the angle between the direction of Earth-to-Sun and
the incoming light ray to the detector on the Earth, and $r_E$ is the Earth-Sun 
distance. The impact parameter is $b=r_E\sin\alpha$. In our case, the 
parameter
$\gamma$ is not a constant. It is given by

\begin{eqnarray}
\gamma = 1- 4 \omega^2 \Delta \left ({4G_N m_\odot\over r_E \sin\alpha}
\right )^2 \ln \left ( {4G_N m_\odot\over r_E \sin \alpha}\right ),
\end{eqnarray}
and depends not only on $\omega^2$, but also
on the angle $\alpha$.

Experimental observations have found no deviations from the EGR theory 
prediction for $\gamma$ from radio waves to
visible light. 
For $M_S = 1$ TeV, the typical limit set by most of the collider experiments,
there is no conflict for photons with frequencies below the visible.
Experimental observations of gravitational lensing by the Sun
in visible light from whole sky survey of Hipparcos 
have found\cite{5} $\gamma = 0.997 \pm 0.003$. 
This is a very impressive result. Unfortunately this value can not be
used directly in our case because in the Hipparcos analysis, $\gamma$
was assumed to be constant in the whole range of $\omega$ and $b$ and
most of the data was at large $b\geq r_E/2$. In our case
the largest deviation from EGR is reached for light grazing the Sun.
In this region the accuracy of the observations 
is not as good as the whole sky result.
The result  for visible light near the solar limb 
is $\gamma = 0.95\pm 0.11$\cite{visible}, which is considerably less 
accurately measured. However even with such accuracy,
we find that the
mass $M_S$ is constrained to be larger than about 
$1.4 (2/(n-2))^{1/4}$ TeV at $2\sigma$
level
for $n>2$ and a factor of approximately 3 larger
for $n=2$. This bound is comparable with the 
limit obtained from collider data.
Radio data from sources near the 
Sun give $\gamma = 1.001\pm 0.002$\cite{radio}, 
consistent with 1 as we would expect for very low frequency photons. 
With $\gamma$-rays of energy one MeV, no observed deviation from EGR theory 
up to 10\% would imply that the scale $M_S$ must be larger than
$1.5\times 10^3$ TeV which is much stronger than any collider 
experimental bounds. 

We suggest that future studies of the parameter $\gamma$ should vigorously
investigate its frequency dependence and its impact parameter dependence.
Theories of the type considered here, with mass $M_S$ about 3 TeV scale,
suggest $\gamma -1$ is negligible for radio frequencies, is positive of order
$3\times 10^{-3}$ in the visible and is so large at $\gamma$-ray 
frequencies that our approximation are no longer valid 
for light grazing the Sun. For larger impact parameters, the effect
can become much smaller.

The same analysis can be carried out for other systems. Due to smaller ratios of
mass to radius for the planets in the solar system, the corrections for the
gravitational lensing by planets in the solar system are small beyond the reach for 
near future experiemnts. However, gravitational lensing by heavier objects, such as
qusars with known masses and radii 
 the effects of the extra dimensions can be large. Precision 
experiments on gravitational lensing for these objects can provide important information
about the theory of gravity and possible extra dimensions.
 
This work was supported in part by the National Science Council of R.O.C under
Grant NSC 88-2112-M-002-041 and by the Australian Research Council. We thank 
Dr. R. Webster for helpful discussions.

\begin{figure}[htb]
\centerline{ \DESepsf(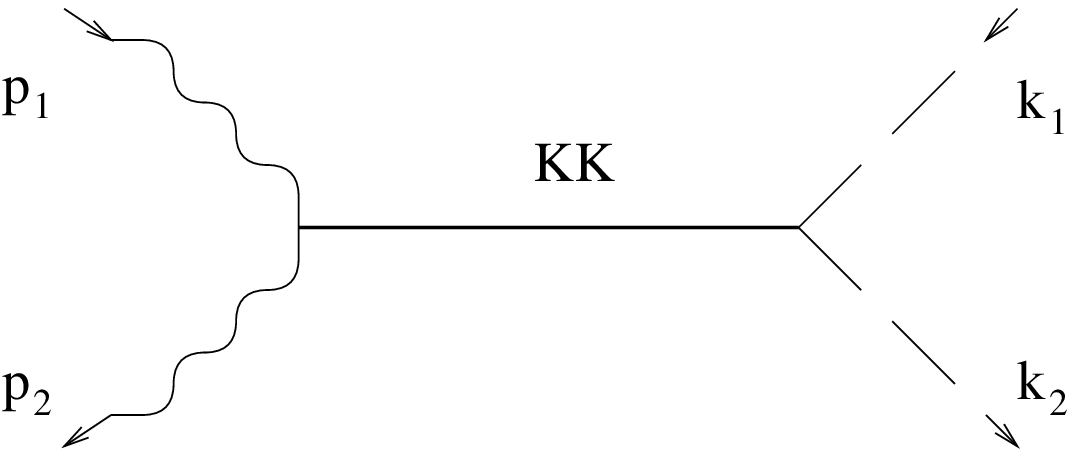 width 10 cm) }
\smallskip
\caption {The Feynman diagram for KK states contribution to
$\gamma(p_1) \gamma(p_2) \to S(k_1) S(k_2)$.}
\label{penguin}
\end{figure}

\end{document}